\documentclass[sigconf, nonacm]{acmart}
\makeatletter
\def\@ACM@checkaffil{
    \if@ACM@instpresent\else
    \ClassWarningNoLine{\@classname}{No institution present for an affiliation}%
    \fi
    \if@ACM@citypresent\else
    \ClassWarningNoLine{\@classname}{No city present for an affiliation}%
    \fi
    \if@ACM@countrypresent\else
        \ClassWarningNoLine{\@classname}{No country present for an affiliation}%
    \fi
}
\makeatother
\usepackage[utf8]{inputenc}
\usepackage{amsmath}

\usepackage{mathtools}
\usepackage{amssymb}
\usepackage{listings}
\usepackage{array}
\usepackage{xspace} 
\usepackage[normalem]{ulem} 
\usepackage{graphicx}
\graphicspath{ {./figures/} }
\usepackage{varwidth}
\makeatletter
\newif\if@restonecol
\makeatother

\usepackage[ruled, linesnumbered]{algorithm2e}
\usepackage{enumitem}
\usepackage{footnote}
\usepackage{tablefootnote}
\usepackage{tabularx}
\usepackage{subcaption}

\setlist{  
  listparindent=\parindent,
  parsep=0pt,
}

\setcopyright{none}

\newif\ifsubmission
\submissionfalse

\ifsubmission

\else

\fi
\newcommand\vldbdoi{XX.XX/XXX.XX}

\newcommand\vldbavailabilityurltwo{https://github.com/denizturkcapar/Quantifying-Uncertainty-Aggregate-Queries}

\begin{document}

\title{Quantifying Uncertainty in Aggregate Queries over Integrated Datasets}

\author{Deniz Turkcapar}
\affiliation{%
  \institution{University of Chicago}
}
\email{dturkcapar@uchicago.edu}

\author{Sanjay Krishnan}
\affiliation{%
  \institution{University of Chicago}
}
\email{skr@cs.uchicago.edu}

\begin{abstract}
Data integration is a notoriously difficult and heuristic-driven process, especially when ground-truth data are not readily available. This paper presents a measure of uncertainty by providing maximal and minimal ranges of a query outcome in two-table, one-to-many data integration workflows. Users can use these query results to guide a search through different matching parameters, similarity metrics, and constraints. Even though there are exponentially many such matchings, we show that in appropriately constrained circumstances that this result range can be calculated in polynomial time with bipartite graph matching. We evaluate this on real-world datasets and synthetic datasets, and find that uncertainty estimates are more robust when a graph-matching based approach is used for data integration. 

\end{abstract}

\maketitle

\section{Introduction}
Data integration is almost always a heuristic process~\cite{halevy2006data}.
Analysts today have no tools to systematically reason about or quantify uncertainties and alert them of potential biases in the integration process. While most statistical models are robust to uncertainty and random errors, the errors introduced by data integration are often unevenly distributed through a dataset. 
\emph{This paper explores how to estimate uncertainty in query results after two datasets are linked by a heuristic entity matching workflow.}
This paper's contribution is a framework that determines the maximal and minimal range of values an aggregate query can take for a given data entity matching workflow.
This is an important contribution to quantifying systematic uncertainty because maximal and minimal ranges can help us understand how uncertainty propagates through results, avoid comparisons between query results with widely varying degrees of uncertainty, and provide confidence when weighting metrics for downstream tasks.
This idea is highly related to an emerging line of work that studies quantifying uncertainty in database aggregation queries over dirty data~\cite{liang2020fast, feng2021efficient}, and we extend these ideas efficiently to a multi-table setting.

It is often the case that the data relevant to a modeling task reside in multiple different datasets. 
For example, one might need to link a weather and a traffic dataset to understand the relationship between weather and automobile accidents. Another common use case is that data scientists might need to combine product catalogs from different business units. 
Independently collected datasets rarely align in terms of format, granularity, and data quality~\cite{magnani2010survey}. 
The integration process is time-consuming, and practitioners have to design highly complex similarity metrics to match corresponding entities or observations~\cite{seligman2002data}. 
This variability would be unseen to most analysts working with the final integrated dataset as the integration process would resolve these candidate sets to a \emph{single match}. However, that latent uncertainty is present in every query result, i.e., there were multiple possible entity matchings that an analyst could have chosen. 

\begin{table}\footnotesize\vspace{-0.25em}
  \label{tab:freq}
  \begin{tabular}{|cc|}
    \toprule
    Product\_Name & Price \\
    \midrule
    StarGazer Premier Pro & 125.00 \\
    StarGazer Academic & 85.00  \\
    ... & ...  \\
    Extended Warranty & 15.00  \\
  \bottomrule
\end{tabular}
\hspace{0.25em}
  \begin{tabular}{|cc|}
    \toprule
    Product\_Tag & N\_Complaints \\
    \midrule
    StarGazer Premier Pro &  33\\
    \textcolor{red}{StarGazer} &  51\\
    StarGazer Academic & 5  \\
    ... &  ...\\
    Extended Warranty & 17\\
  \bottomrule
\end{tabular}
\vspace{0.5em}
\caption{Uncertainty is often unevenly distributed through integrated dataset where some entities (``Extended Warranty'') have clear correspondences, and others (``StarGazer'') have multiple viable matches. \label{table:teaser} }
\vspace{-3.5em}
\end{table}

To make this concern more concrete, suppose you are a data scientist at a software company trying to understand how the price of a product relates to the received customer service complaints. 
The data shown in Table \ref{table:teaser} illustrates two datasets: a product catalog and a customer service log. 
The customer service log is populated by tags manually provided by customer service representatives. As a result, those tags do not always align with names in the product catalog.
Your company's flagship product (``StarGazer'') has many different editions, and the edition name is sometimes missed by the service representatives (leading to the tag marked in red).
To integrate the two datasets, you have one of three choices: either assign the ambiguous tags to one of the ``StarGazer'' products, somehow divide them among the relevant products, or drop those observations from the dataset altogether. 
Each of these choices introduce a different bias in the ``StarGazer'' complaint field, either erring on the side of counting low or high.
The data integration engineer making this decision may not know how the data will be used, and the modeling errors that might arise from such a choice.
Such a scenario is far from rare, and a number of recent papers have highlighted organizational challenges in data integration~\cite{haas2015wisteria, fernandez2018aurum, halevy2006data}.
In our example, some entities (``Extended Warranty'') have clear correspondences with their product tags, but others (``StarGazer'') have multiple viable matches.
Our example dataset is designed to be uncontroversial, but one could imagine how such issues may introduce unintentional bias or equity issues in more consequential datasets~\cite{mehrabi2021survey}.

Data integration is a broad area, and we focus on a narrow sub-problem of two-table similarity matching problems, which is common in a number of data science tasks, in order to evaluate how uncertainty propagates in the query result.
In our two table problem, one of the tables is a \emph{base table}, and one of the tables is an \emph{augmenting table}.
Rows in the \emph{base table} correspond to unique entities or observations, e.g., the product catalog in Table \ref{table:teaser}.
Rows in the \emph{augmenting table} link to at most one ground-truth row in the base table, e.g., the customer complaint table in Table \ref{table:teaser}.
Thus, there is a one-to-many relationship between rows in the base table and the rows in the augmenting table.
However, the exact linkage between the \emph{base table} and the \emph{augmenting table} is not known in advance.
This structure mirrors data integration problems commonly seen in machine learning and data science.

Our framework requires the following inputs: a base and an augmenting table pair, a similarity metric that identifies candidate matches between the two tables, a maximum number of expected matches (corresponds to n variable in 1-n matching), and a SQL aggregate query. 
The framework calculates the minimum and maximum value the SQL aggregate can take over all valid resolutions, i.e., matchings that end in a one-to-many (1-n) relationship.
Let's consider the example in Table \ref{table:teaser}.
Imagine a similarity metric that uses a Jaccard or Overlap similarity metric to match the tables. The ``StarGazer'' row in the complaints table can plausibly match with any of ``StarGazer Premier Pro'', ``StarGazer Premier'', and ``StarGazer Academic''. 
Intuitively, in the integrated table with all products and complaints, the 51 complaints from the ``StarGazer'' row could add to any of the three possible matches. 
\begin{lstlisting}
>>> SELECT SUM(N_Complaints) 
    FROM Integrated 
    WHERE Product_Name = 'StarGazer Premier Pro'
    
result: 33 <= SUM_N_Complaints <= 51 + 33 = 84
\end{lstlisting}
Our framework finds these maximal and minimal ranges (84 and 33 respectively in this example) given a user inputted estimation of maximum number of matches (n). 

In short, this paper makes the following contributions.
\begin{itemize}
    \item We propose a framework for estimating the variation in a query result due to matching choices. \emph{Unlike precision and recall, this metric does not require ground truth data}. 
    \item We propose an algorithmic framework based on graph matching to efficiently calculate this uncertainty measure for different SQL predicates and aggregation functions of interest.
    \item We illustrate how these downstream query result uncertainty metrics can be used to inform downstream data science applications through a real-world application of our framework.
\end{itemize}

\section{Background}
\subsection{Related Work}
The ideas in this paper are highly related to the concept of ``reverse data management'' proposed by Meliou et al.~\citep{meliou2011reverse, meliou2012tiresias}. 
Meliou et al. argue that as data grow in complexity, analysts will increasingly want to know not what their data currently says but what changes have to happen to the dataset to force a certain outcome.
Such \emph{how-to} analyses are useful in debugging, understanding sensitivity, as well as planning for future data.
Meliou et al. build on a long line of \emph{what-if} analysis and data provenance research, which study simulating hypothetical updates to a database and understanding how query results might change~~\citep{deutch2013caravan, buneman2001and}.

We find that there is a gap in the literature when it comes to data integration problems, which are generally not easily expressible in standard relational algebra~\cite{halevy2006data}. One decade ago, there was some interest in uncertainty management for data integration~\cite{magnani2010survey}. While these papers have been successful in the confines of inference and approximation, they do not concern themselves with quantifying the uncertainty propogated to downstream query results. Similar to~\cite{liang2020fast}, we find that quantifying extremal behavior of aggregate queries is substantially more objective than parameterizing an entire probability distribution.

In the confines of data integration, there has been considerable efforts in entity matching techniques~\cite{entity-res-dataset-origin-2010}\cite{moma-dataset-ref}\cite{dedoop-dataset-ref}\cite{learning-based-dataset-ref}. However, these papers also do not consider ways in which we can quantify uncertainty in query outcomes of integrated datasets. Our paper makes use of the datasets that these papers have used, and we expand on the idea of how we can create hard bounds for quantified uncertainty in the downstream query result. 

This unevenness of uncertainty is an instance of an often under-appreciated statistical phenomenon called \textbf{heteroscedasticity}.
Formally, heteroscedasticity happens when the variability of the random disturbance is different across elements of the vector.
The database equivalent is where different rows (or more generally different query predicates) have wildly different levels of certainty of their values. Kaufman suggests that about twice as many articles should be testing and correcting for data heteroscedasticity than they currently do ~\citep{kaufman2013heteroskedasticity}. Kaufman calls for action to care more about heteroscedasticity by pointing out that it is more common than it is usually recognized by researchers. 

\subsection{Problem Statement and Preliminaries}
Now, we will introduce a formal problem statement that will guide the remainder of the technical discussion.
Every table has a set of identifying attributes and measurement attributes.
Identifying attributes describe the entity a row refers to, and measurement attributes describe a quantifiable property of the entity.
For example, let $R$ be a relation over the attributes $A$, the attribute set can be decomposed into $R[A_{id} \cup A_{measurement}]$.
If $\mathcal{U}$ denotes a universe of real-world entities, there exists a mapping between each row to a corresponding real-world entity through the identifying attribute:
\[
C: \Pi_{A_{id}}(R) \mapsto \mathcal{U},
\]
where $\Pi$ denotes the standard projection operator.
In our example dataset, the real-world entities are actual products that the company sells. The \texttt{Product\_Tag} is an identifying attribute that corresponds to some real-world entity, and the \texttt{N\_Complaints} attribute is a measurement.

\subsubsection{Base and Augmenting Relations}
A table is called a ``base relation'' if the mapping of each row in $R$ represents information about a unique real-world entity.

\begin{definition}[Base Relation]
Let $R$ be a relation and $r[A_{id}]$ denote the projection of a row of $R$ onto the identifying attributes. $R$ is a base relation if two conditions hold: 

\noindent (1) For two rows $r,s$, if $r[A_{id}] = s[A_{id}]$ then $r = s$.

\noindent (2) For two rows $r,s$, if $C(r[A_{id}]) = C(s[A_{id}])$ then $r = s$.
\end{definition}

Primary relations are tables that have already been deduplicated (or do not have to be).
For example, a master employee table in a company or patient database in a hospital.
Such a table is like our \texttt{(Product\_Name, Price)} table in our intro example.
It is important to have at least one such table when studying this problem, because we need a clear unit for uncertainty measurement, i.e., a precise notion of an individual from a statistical population.

Any table that is not a base table is called an \emph{augmenting table}. Base tables can be linked to one or more augmenting tables when these tables refer to the same real-world entities.
However, as standard in data integration problem, we assume that this mapping is not trivial.
That is, the augmenting table has a different correspondence with real-world entities than the base table.
So, an entity matching procedure must be applied to combine the base table and augmenting tables.

In notation, let $S$ be an augmenting table over the attributes $B = B_{id} \cup B_{measurement}$. Rows $S$ corresponds to a property, measurement, or some other form of information about a real-world entity.

Let $R[A]$ be a base relation and $A_{id}$ denote its identifying attributes. 
Let $S[B]$ be an augmenting relation and $B_{id}$ denote its identifying attributes.
\begin{definition}[Valid Augmentation]
A \emph{valid} augmentation is a procedure that returns a relation $M[A,B]$ over both sets such that the following integrity constraints hold. A functional dependency:
\[M[B_{id}] \rightarrow M[A_{id}]\]
and an inclusion dependency: 
\[
S[B_{id}] \subseteq M[B_{id}]
\]
\end{definition}
In other words, we want a consistent mapping between augmenting table entities and base table entities. That consistency condition can be described in terms of a functional dependency. We also want this mapping to capture all of the information in $S$, which is described by the inclusion dependency.

\subsubsection{Entity Matching Model}
As described above, augmentation can be the result of nearly any computational procedure. We now make that a bit more precise to understand the uncertainties in this matching process.
Most record-linkage procedures operate in two steps: (Step 1) candidate set of possible matches are produced, (Step 2) the candidate set is resolved to certain matches. We argue that the uncertainty by Step 2 is often ignored. One can think of Step 2 as selecting one possible ``matching scenario'' out of combinatorially many that could be allowed by the candidate sets. 
In this paper, we explore the following technical question: Given a description of Step 1, can we determine just how much uncertainty is ignored.

Let $\Psi(r,s) \subseteq R \times S$ be a relation that defines the candidate matching set. This is a subset of the Cartesian product of $R \times S$ and it describes all possible linkages between the described entities.
Contained in $\Psi$ are a number of valid augmentations. This set can be denoted as:
\[
\mathcal{M} = \{M: M \subseteq \Psi ~~,~~ M \text{ is valid}\}
\]
We assume that $\mathcal{M}$ contains $M^*$ the true matching between identifying attributes in $S$ and $R$.

With this problem in mind, we consider hypothetical SQL queries of the following form over the integrated data:
\begin{lstlisting}
SELECT {COUNT,SUM, AVG}(measurement)
FROM Integrated 
WHERE Predicate(a1,...,aN,b1,..,bL)
\end{lstlisting}
Where the aggregate is over any of the measurement attributes, and the predicate is Boolean condition over any of the attributes in either table. 
We can denote such a query as $q(M)$ applied to some integrated relation.
Our objective is to produce the following bounds:
\[
u = \max_{M \in \mathcal{M}} q(M) ~~~,~~~ l = \min_{M \in \mathcal{M}} q(M)
\]
Under our assumptions, this means that $q(M^*)$ is contained in the range $[l,u]$.
We call the tuple $[l,u]$ the \emph{result interval}.

\begin{definition}[Problem Statement]
Given a base relation $R$ and an augmenting relation $S$, and a $\Psi$ that defines candidate matches between $R$ and $S$, find the maximal value $u$ and minimal value $l$ that an aggregation query can take over all valid augmentations defined by $\Psi$.
\label{def:prob}
\end{definition}

Note that our careful definition of valid augmentation is what makes this a well-posed problem. Without the inclusion dependency $l$ is trivially 0 for all queries (assuming positive valued data), because an empty set would be a valid augmentation. We will see how this definition factors in to the eventual optimization problem that we solve to compute $l$ and $u$.

\subsection{The Naive Solution}
Existing frameworks have made initial progress towards such problems problem~\cite{potti2015daq, liang2020fast}, but they are limited in how they handle queries across multiple tables. DAQ~\cite{potti2015daq} defines a concept of upper and lower relations to bound the result of uncertain aggregates.
Associated with each cell in a table is a range of possible values, and these ranges can be propagated through aggregate queries.
One could use an approach inspired by DAQ to estimate $l$ and $u$.

For example, first we would construct the following sets for each row $r$ of the base table:
\[
M_r = \{s: (r,s) \in \Psi\}
\]
This is the set of $s$ rows that match with the chosen $r$.
For every attribute in $b \in B$, one can calculate:
\[
u_b = \max M_r.b ~~,~~ l_b = \min M_r.b
\]
for each $r$, and assume that categorical attributes are dictionary encoded.
If we did this for each row $r$ this would create table over attributes $A$ and $B$ with value ranges on each $B$ value.
Using an algorithm like the one in~\cite{potti2015daq}, one can compute an upper bound and lower bound for all SUM,COUNT, and AVG queries with predicates.

\subsubsection{Example}
We will show that this basic technique can lead to highly misleading results in even simple cases. 
For example, consider our example dataset, and a $\Psi(r,s)$ that identifies pairs of rows where the \texttt{Product\_Name} and \texttt{Product\_Tag} have a Jaccard similarity of greater than 0.3.
This metric would map  the ``StarGazer'' \texttt{Product\_Tag} to the \texttt{Product\_Name} ``StarGazer Academic'' and ``StarGazer Premier Pro''.
\begin{table}[ht!]\tiny
  \label{tab:freq}
  \begin{tabular}{|cc|cc|}
    \toprule
    Product\_Name & Price & Product\_Tag & N\_Complaints \\
    \midrule
    StarGazer Premier Pro & 125.00 & \{StarGazer Premier Pro, \textcolor{red}{StarGazer}\} &  \{33, 51\}\\
    StarGazer Academic & 85.00  &
    \{StarGazer Academic, \textcolor{red}{StarGazer}\} & \{5,51\} \\
    Extended Warranty & 15.00 &
    Extended Warranty & 17\\
  \bottomrule
\end{tabular}
\end{table}
If we were to use~\cite{potti2015daq} to evaluate a total number of complaints on all products, the maximum possible value would be 119 and the minimum would be 55.
Both of these values are misleading. 
In the maximum, we would double count the complaints for the ``StarGazer'' \texttt{Product\_Tag}, and in the minimum none of the ``StarGazer'' counts are included. In a sense, the uncertainty in these values is coupled, where setting one of the values affects what the other could be. Furthermore, this strategy is highly susceptible to outliers. Going back to the real data with the Google-Amazon product matching, we can see that real-world similarity metrics can be very imprecise for some matches. Even one erroneous match could completely skew an uncertainty estimate. 
\section{A Graph Matching Approach}
The final example in the last section illustrates two key problems in assessing uncertainty in integrated datasets: \textbf{Coupling} and \textbf{Robustness}.
The \textbf{coupling} problem is when the value for one row depends on the choice of value for another.
The \textbf{robustness} problem is to mitigate the affect of outliers that can affect upper and lower bound calculations.
We show that both issues can be elegantly solved with an optimization problem called unbalanced assignment (a generalization of bipartite matching).

\subsection{The Coupling Problem}
The set $\Psi(r,s)$ can be thought of as defining a graph over the entities in $R$ and $S$. 
Let $V_R, V_S$ be defined as the set of all identifying tuples from both $R$ and $S$ respectively.
\[
V_R = \Pi_{A_{id}}(R)~~~,~~~ V_S = \Pi_{B_{id}}(S)
\]
We can define a bipartite graph between these sets where an edge exists if there exists an $r[A_{id}],s[B_{id}] \in \Psi(r,s)$.

Every valid augmentation can described as a subgraph of this bipartite graph where each $v_s \in V_S$ has an edge to at most one $v_r \in V_R$. Such a subgraph is called an unbalanced assignment, i.e., assigning a $v_s$ to a $v_r$.
The existence of such subgraphs gets to the essence of the coupling problem shown by the example in the previous section.
If we match one pair $r[A_{id}], s[B_{id}]$ of identifying tuples, it affects how we can match others.
Double counting happens because we don't appropriately account for this.

\subsubsection{Finding Optimal Generalized Assignments}
It turns out that our problem of uncertainty quantification reduces to finding extremal generalized assignments in a bipartite graph. Let's see how this works in the abstract first.
Let $W(r,s)$ define weights on each edge of the graph above, and let $x_{r,s}$ denote an indicator function if an edge is kept or removed.
We can define two optimization problems. The first problem finds an unbalanced assignment with the highest cumulative edge weights:
\begin{equation}
\max_{\mathbf{x}} \sum_{(r,s) \in \Psi} W(r,s) \cdot x_{r,s}
\label{eq:max}
\end{equation}
\[
\text{subject to: } \forall s \in S: \sum_{r \in R} x_{r,s} \le 1
\]
\[
x_{r,s} \in {0,1}
\]
\[
\forall r,s \not \in \Psi x_{r,s} = 0
\]
We can see that even though the constraint is $x_{r,s} \le 1$, it would be attained by making as many values $1$ as possible. The last constraint is a technicality to allow for ``0'' edge weights and to differentiate between a 0 weight and a non-existent edge.

And, we can also consider the minimization version of this problem.
\begin{equation}
\min_{\mathbf{x}} \sum_{(r,s) \in \Psi} W(r,s) \cdot x_{r,s}
\label{eq:min}
\end{equation}
\[
\text{subject to: } \forall s \in S: \sum_{r \in R} x_{r,s} \ge 1
\]
\[
x_{r,s} \in {0,1}
\]
\[
\forall r,s \not \in \Psi x_{r,s} = 0
\]
We can see that even though the constraint is $x_{r,s} \ge 1$, it would be attained at equality 1 since it is a minimization. Thus, both directions of this problem result in valid matchings.
We will show in the next section that result interval calculation essentially boils down to appropriately setting $W(r,s)$ for different query types. However, before we discuss that, let's discuss how such problems are solved.

At first glance, this problem seems like a Mixed Integer Linear Program, but it can actually be efficiently solved. Notice that it is a generalization of a bipartite matching problem. 
In that case, the left and right parts of the bipartite graph constructed from sets A (left) and B (right) and one is looking for a one-to-one match. Then, the Blossom algorithm, also called Edmonds’ matching algorithm, is applied to end up with a matching that improves along augmenting paths, which involves the process of alternating paths between unmatched vertices. The Blossom algorithm is a generalized version of the Hungarian algorithm that can be used on any graph to construct a maximum matching. The Blossom algorithm improves on the Hungarian matching algortihm by shrinking cycles in the graph to reveal augmenting paths when constructing a maximum or minimum match. The overall complexity of the Blossom approach for bipartite matching is $O(n^3)$, where $n$ is the number of vertices.	
Now, consider the unbalanced assignment problem where matches are no longer one-to-one. 
The basic idea is straightforward. 
One simply creates dummy vertices to balance the problem and reduce it to bipartite matching.
Let $d$ be the maximum in-degree of a vertex in $V_r$. 
Every vertex in $V_r$ is duplicated $d$ times and each new vertex has the same edges as its duplicate.

A one-to-one matching over this new graph produces an assignment.
This problem is essentially reduced to the aforementioned balanced assignment problem by adding b - a new vertices to the left part of the graph constructed using elements from set A and connecting them with the members of the right side.  After this process, the Blossom algorithm is applied to end up with matchings. This algorithm has an $O(d^3 n^3)$ complexity. 

\subsection{Interpreting Results}
We denote the maximization problem $GA_{max}$ and the minimization problem $GA_{min}$. These core subroutines in our result interval estimation algorithm. Let's now try to understand when these optimization problems are meaningful. Suppose, we have a candidate set $\Psi$ and contained in this candidate set is the correct matching $M^*$ between the tables $R$ and $S$. $M^*$ can be thought of as subgraph of $\Psi$. For any weighting function, let $\mathcal{W}(M^*)$ be:
\[
\mathcal{W}(M^*) = \sum_{(r,s) \in M^*} W(r,s) 
\]
It follows that:
\begin{proposition}
If $M^*$ is contained in $\Psi$, then for any weighting function $\mathcal{W}(M^*)$ is upper bounded by the solution to $GA_{max}$ and lower bounded by the solution to $GA_{min}$.
\end{proposition}

This proposition gives us an understanding of when it is possible to bound the range of values a sum of weights could take. Namely, it is only possible if the true matching is contained in the candidate set. This is less a statement of assumption and more a statement of semantics:

\vspace{0.25em}

\noindent \emph{Given any candidate set, our framework returns the range of valid aggregate results over all matchings contained in the candidate set.}

\vspace{0.25em}

This proposition is true for all weighting functions and we will show the an individual query processing instance can be represented as a specific weighting function. 

\subsection{Robustness Problem}\label{sec:constraint}
The asymmetry between false positives and false negatives creates another issue. 
Notice that the primary constraint is of the form $\forall s \in S: \sum_{r \in R} x_{r,s}$. In a sense, this constrains the number of $s$'s that can match with an $r$ but not the reverse. 
While this is notionally true based on our definitions, it can lead to serious robustness issues.

Consider the real data from the Amazon-Google product matching dataset. While the candidate set matches most products nearly one-to-one, some products match with nearly 120 others.
The imprecision in the candidate set can create implausible assignment scenarios where hundreds of different $S$ entities match with a single one in $R$.
Similar to the concept of regularization in machine learning, we need to constrain the optimization problem to penalize degenerate solutions that might occur in highly skewed candidate sets.
We add a constraint of the following form that limits the in-degree of each $r$ in final the assignment to a value $N$:
\[
\forall r \in R: \sum_{r \in R} x_{r,s} \le N
\]

Note that the addition of this constraint may make the optimal matching ``invalid'' as some of the entities in $S$ may not be matched with entities in $R$. But, it is a crucial knob for the user to control the resilience to skew. One can think of $N$ as analogous to the probability in a statistical confidence interval: the higher the probability the more conservative the intervals are. $N$ serves a similar function.

\subsubsection{Solving the Constrained Problem}
Interestingly enough, adding this constraint does not change the hardness of the problem.
We can make a small tweak to the general algorithm for unbalanced assignments.
Rather than duplicating each $V_r$ $d$ times, we only duplicate it $N$ times.
Not only does this control the outliers, it also significantly improves the time-complexity by reducing the dependence on a large $d$.

The entire workflow is described in Algorithm 1.
Note that the algorithm creates a weighted graph and solves the assignment problem over this weighted graph. It should also be noted that a maximum cost assignment problem could be converted to solving a minimum cost assignment problem, and vice versa~\cite{goeman}. Assume that the cost matrix is $c$, solving the maximum cost assignment problem for $c$ is equivalent to solving the minimum cost assignment problem for cost matrix $-c$. In order to do the weight transformation, we need to solve $MaxWeight(Graph) - w_i$, and assign the outcome value to that edge, where $w_i$ indicates individual edges. Blossom algorithm gives a maximal matching, but because we can use the property that min and max matching problems can be transformed, we can use a transformed min weighted graph to solve minimum matching problem using Blossom technique, which is a generalized version of the Hungarian Algorithm. So we will transform the max matching graph in such way that the weights reflect negative cost, and this way we can apply the Blossom maximum matching algorithm, which works by running the Hungarian algorithm when the graph is completely bipartite - which is true for our case -, to the graph~\cite{kolmogorov2009blossom}. Even though we used the Blossom maximum matching technique, the Hungarian maximum matching algorithm can also be used to solve the problem and it would give the same outcome.
The particular choice of weights is query specific and will be described in the next section.

\SetKwComment{Comment}{/* }{ */}
\begin{algorithm}[t]
\KwData{2 tables ($R$ and $S$), $N \geq 0$ number of max matching constraint, a candidate set $\Psi$, and a weighting function $W(r,s)$.}
\KwResult{Maximum and Minimum Matching Generalized Assignment Solution}
\caption{Constrained Maximum and Minimum Matching Problem Using Graph Approach}\label{alg:cap}
$G \gets Graph$\;
$R = R.\texttt{dup(N)}$ \Comment*[r]{N copies of each row in R}
$MaxM \gets Set$ \Comment*[r]{Maximum Matching Set}
$MinM \gets Set$ \Comment*[r]{Minimum Matching Set}
\For{\texttt{$r \in R$}}{
    \For{\texttt{$s \in S$}}{
         \uIf{$(r,s) \in Psi$}
         {
            G.\texttt{addEdge}(r,s,W(r,s))
        } \uElse{
            G.\texttt{addEdge}(r,s,W = 0)
        }
    }
}
$MaxM \gets \texttt{maximumMatchingAlgorithm(G)}$ \;
$maxWeight \gets $\texttt{findMaxWeightValue(G)} \;
$MinGraph \gets Graph$.\texttt{copy()} \;
\For{$u, \ v, \ weight \in MinGraph$}{
    $newWeight \gets maxWeight - weight$
    MinGraph.\texttt{updateWeight(u,v,newWeight)}
}
$MinM \gets \texttt{maximumMatchingAlgorithm(minGraph)}$
\textbf{return} $MinM$, $MaxM$
\end{algorithm}

\subsection{Remarks on the Optimization Problem}
The lines that we have drawn restricting this problem space are crucial. In fact, even slightly more complex versions of this problem become computationally hard.
The key issue is that while bipartite matching (and its constrained variants in this paper) can be solved in polynomial time, tripartite matching is NP-Complete ~\cite{karp1972reducibility}.
For example, if one were to relax the problem statement to allow for base tables that are not de-duplicated such a matching could arise.
In this case, we would have to enforce that that not only is there a functional dependency between the entities in $S$ and $R$, but also within $R$ the matchings are consistent to rows that refer to the same entity. Solving the joint problem is NP-Complete, and the most reasonable heuristic is to first de-duplicate $R$ and then proceed with the interval calculations.


\section{Query Processing}
Next, we will discuss how to use this graph solution to construct result intervals for SQL aggregate queries over the integrated dataset.
As a technical assumption, we will assume that all of the numbers that we are working with are positive.
This is easy to achieve in practice because if they are not, one can simply shift the data so that they are.

\subsection{Setup and Preprocessing}
In our framework, users define candidate matches with a simple API: a similarity measure between entities in $R$ and $S$, and a threshold. This API is sufficient to generate a set of candidate matches $\Psi$. 
Some entity matching frameworks leverage ``blocking'', which subdivides a dataset into blocks before similarity comparisons. This information is also easy to integrate into the framework and would simply remove edges from $\Psi$.
Finally, the user specifies a $N$ which is the maximum number of rows in the augmenting table that could match with rows in the base table.

\subsubsection{Grouping Identifying Attributes}
A hidden subtlety with the algorithm proposed in the previous section is that it operates over distinct identifying attributes and not rows. It is possible that in the table $S$ there are multiple rows with the same identifying attributes (based on our definitions this is not possible in $R$).
As a pre-processing step, we have to group these rows together and treat them as a single matching entity.
For each distinct identifying tuple $S[B_{id}]$, we select a set of rows with that identifying tuple which we denote as $\textbf{s}$ (making it clear that it is possibly a group of rows).
And recall from the last section, that 
$\Psi$ contains a candidate match between entities if ANY of this set matches with $R$ based on the similarity metric.
This is a careful choice of definition to avoid inconsistencies where different rows with the same identifying attribute are matched differently.

\subsection{SUM and COUNT}
First, we will construct the bounds for SUM and COUNT queries.
Once the candidate set of matches $\Psi$ is constructed and the constraint $N$ is known, result interval calculation for SUM/COUNT queries is relatively straight-forward. We simply define the graph weights from the previous section based on the query.

Let $r$ be a row in $R$ and $\textbf{s}$ be a grouped set of rows with the same identifying attribute in $S$. 
We can think of this pairing as a hypothetical sub-table with rows:
\[r\times\textbf{s} = (r,s_0),(r,s_1),...\]
Over this sub-table, we can apply the user-specified SQL predicate, which will return a subset of those pairings $pred(r\times\textbf{s})$.
Let's define the following quantities:
\begin{itemize}
    \item $\text{sum}(r,\textbf{s})$ The total SUM of the SQL aggregation attribute over $pred(r\times\textbf{s})$.
    \item $\text{count}(r,\textbf{s})$ The total COUNT of the SQL aggregation attribute over $pred(r\times\textbf{s})$.
\end{itemize}
For COUNT queries, the weighting $W(r,\textbf{s})$ is .
\[
W(r,\textbf{s}) = \text{count}(r,\textbf{s})
\]
For SUM queries, the weighting $W(r,\textbf{s})$ is defined as:
\[
W(r,\textbf{s}) = \text{sum}(r,\textbf{s})
\]

The solution to the equations \ref{eq:max} and \ref{eq:min} calculate upper and lower bounds for SUM and COUNT queries.
This can be directly seen from the formulas in the equations.
The objective function optimizes a sum over edge weights, and both SUM/COUNT queries can be expressed as a sum. 
Thus, the result finds the minimum and maximum sum of weights over all unbalanced matchings. This is equivalent to finding the minimum/maximum value of SUM/COUNT queries over all valid augmentations contained in the candidate set.

Furthermore, since this section describes how to deal with the multiplicity in $S$ rows, we'll remove the boldface notation $\textbf{s}$ for brevity in the following discussion (with the assumption that it is handled in the same way). 

\subsection{Result Intervals for AVG}
Answering AVG queries are a little more complicated since they are not neatly expressed as sums of weights, with the objective being:
\begin{equation}
\frac{\sum_{(r,s) \in \Psi} W(r,s) \cdot x_{r,s}}{\sum_{(r,s) \in \Psi} \textbf{sgn}(W(r,s)) \cdot x_{r,s}}
\label{eq:avg}
\end{equation}
where $sgn$ is the sign function.
We use a simple trick to estimate average queries by upper and lower bounding this objective. 
Essentially, we consider calculating the result intervals for an equivalent SUM query, we bound the maximum and minimum value that the denominator could take at attainment.

\begin{proposition}
Given an AVG query, let $l_{sum}$ and $u_{sum}$ be the lower and upper bounds for a SUM query with the same predicate. The AVG query is bounded by:
\[
l_{avg} = \frac{1}{\min\{|R|N,|S|\}} \cdot l_{sum}
\]
and,
\[
u_{avg} = \frac{1}{d} \cdot u_{sum}
\]
where $d = \sum_{s \in S} (\max_{r \in R} \textbf{sgn}(W(r,s)))$.
\end{proposition}
The proof of this proposition is contained in the appendix.
This proposition shows that the AVG query can be answered with a scaled version of the SUM query.
Thus, experimentally, we focus our efforts SUM and COUNT queries knowing that AVG queries are essentially the same.

\subsection{Relative Uncertainty}
Beyond their absolute interpretations, result intervals are also useful in relative terms. One can compare result intervals from different query predicates to understand how uncertainty is distributed through a dataset.
Suppose we have a table $\hat{M}$ which is the final valid augmentation that a user chooses, and let $q(\hat{M})$ be a \emph{nominal query result} that is the result of an aggregate query over the final matching. Let $(l,u)$ denote a result interval calculated using the procedures mentioned in the paper.

The relative error in a query result is $rel(q) = \frac{u - l}{q(\hat{M})}$.
This metric normalizes for the fact that some queries have naturally higher results than others.
Using this metric, different queries can be compared in terms of their relative errors, e.g., $rel(q_1) > rel(q_2)$.
This is an instrumental tool to understanding potential systematic biases that might arise from data integration.

\subsection{Numerical Example}
To show how all of this estimation works, we will work through a simple but illustrative numerical example. Let's consider the dataset described in the introduction. We construct a set of candidate matches using a Jaccard Similarity threshold of 0.3. This results in graph with the following adjacency matrix, where rows are base table entities and columns are augmenting table entities.

\begin{table}[ht!]\tiny
  \begin{tabular}{|c|cccc|}
    \toprule
     & StarGazer Premier Pro & StarGazer  & StarGazer Academic & Extended Warranty\\
    \midrule
    StarGazer Premier Pro & 1 & 1 & 0 & 0 \\
    StarGazer Academic & 0 & 1 & 1 & 0  \\
    Extended Warranty & 0 & 0 & 0 & 1  \\
  \bottomrule
\end{tabular}
\end{table}

Queries are processed as weighted graphs over such a matrix. Consider the query:

\begin{lstlisting}
>>> SELECT SUM(N_Complaints) 
    FROM Integrated 
    WHERE Product_Name = 'StarGazer Premier Pro'
\end{lstlisting}

This query would generate the following weighted graph, described as its adjacency matrix:
\begin{table}[ht!]\tiny
  \begin{tabular}{|c|cccc|}
    \toprule
     & StarGazer Premier Pro & StarGazer  & StarGazer Academic & Extended Warranty\\
    \midrule
    StarGazer Premier Pro & 33 & 51 & 0 & 0 \\
    StarGazer Academic & 0 & 0 & 0 & 0  \\
    Extended Warranty & 0 & 0 & 0 & 0  \\
  \bottomrule
\end{tabular}
\end{table}

A valid augmentation assigns all $S$ entities to only one $R$ entity. We can easily see that the following assignment is a maximizer: 'StarGazer Premier Pro -> StarGazer Premier Pro', 'StarGazer -> StarGazer Premier Pro', 'StarGazer Academic -> StarGazer Academic', 'Extended Warranty -> Extended Warranty'.
The maximizer has a SUM of 84.
Likewise, we can see that the minimizer is an assignment: 'StarGazer Premier Pro -> StarGazer Premier Pro', 'StarGazer -> StarGazer Academic', 'StarGazer Academic -> StarGazer Academic', 'Extended Warranty -> Extended Warranty'.
The minimizer is a valid augmentation because every $S$ is assigned along an edge in the original adjacency matrix.
The minimizer has a value of $33$.

\section{Experiments}
Next, we present a series of experiments on real and synthetic data to show how the result interval estimation works. 

\emph{Objectives. } Our objective is to accurately estimate the range of values an aggregate can take over different possible linkages of two tables. In all of our experiments, we consider a ground truth matching of the two tables to calculate a baseline aggregate query result. We compare this baseline query result to the estimated intervals on the following axes. 

\begin{enumerate}
    \item (Tightness) How tightly do the calculated results bound a ground truth aggregate value in real data?
    \item (Failure Rate) How often does the ground lie outside the calculated range?
    \item (Calibration) How well does the length of the result intervals correlate with more and less uncertainty in the linkage?
\end{enumerate}

\subsection{Real-Life Datasets and Setup}
We consider real-life datasets, which have a ground truth matching, in our experiments so that we can compare actual numbers to our interval calculations ~\footnote{Please see following for access to used datasets: https://dbs.uni-leipzig.de/research/projects/object\_matching/benchmark\_datasets\_for\_entity\_resolution}. All the datasets are ``title'' matching datasets like our example, and we use a Jaccard similarity metric. It should be noted that the similarity metric can be any similarity metric provided by the user because the specific similarity metric is not the point of our experiments. For our purposes, Jaccard similarity is an appropriate similarity metric for all real data sources that have been used. The table below shows statistics information relevant for our investigation about the real datasets.

\begin{figure*}[t]
    \centering
    \includegraphics[width=0.25\textwidth]{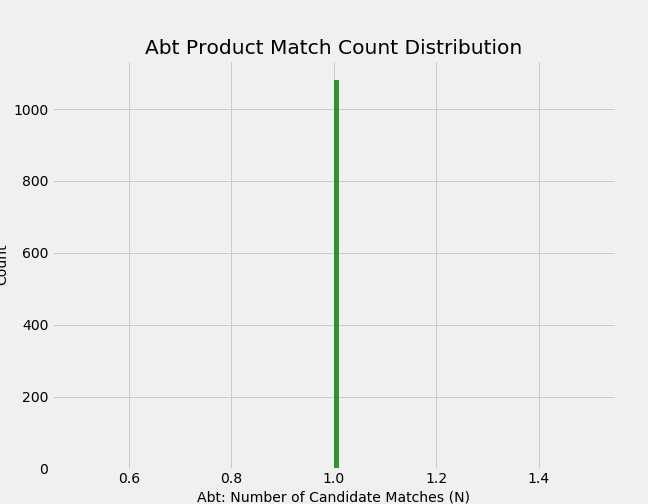}
    \includegraphics[width=0.25\textwidth]{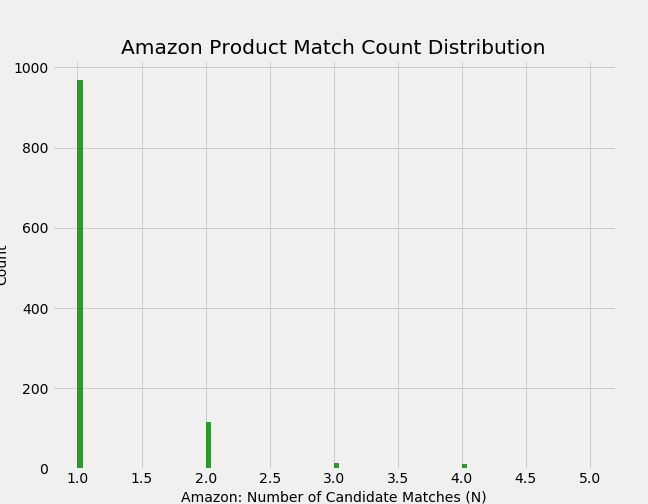}
    \includegraphics[width=0.25\textwidth]{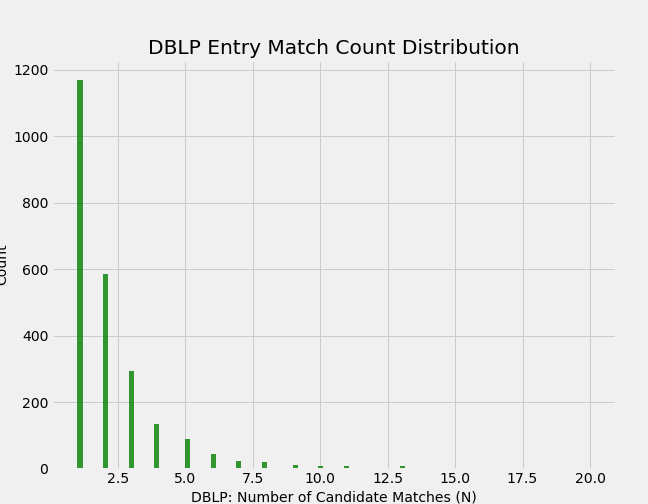}
    \caption{Distribution of ground truth matches in the datasets.}
    \label{fig:datastats}
\end{figure*}

\subsection{Baseline Algorithms}
The following algorithms give us alternative result interval estimates. 

\vspace{0.5em} \noindent \textbf{Max-Sum: } This approach takes every $r$ in the base table, and considers every possible match in $S$ to construct a range of possible values for the aggregation attribute. 
The upper and lower bounds are constructed with an extension of the techniques in~\cite{potti2015daq, liang2020fast}. It multiplies the total number of matches by the the maximum of the range to derive an upper bound. It simply takes the lower of the range to construct a lower bound.
We call this baseline the ``Max-Sum'' baseline because it relies on the max-sum inequality to construct a result range. If the candidate set contains the ground truth, this baseline is guaranteed to bound the true result. However, the minimum and the maximum values are not tight because they are not necessarily valid matchings based on our problem statement.

\vspace{0.5em} \noindent \textbf{Max-Sum+C: } The Max-Sum baseline is highly susceptible to outliers (due to the maximum value multiplied by the cardinality). We next consider a version of the Max-Sum baseline that is constrained. This approach clips the the upper bound by using a constraint of the maximum number of matches that could (exactly like Section \ref{sec:constraint}). We use a heuristic to determine the appropriate constraint which is the 75\% percentile of the number of matches of every $r$ in the candidate set. This baseline is no longer guaranteed to bound the true result but is often much tighter than the naive Max-Sum bound. 

\vspace{0.5em} \noindent \textbf{Generalized Assignment (GA): } Our proposed algorithm runs without a matching cardinality constraint. Like the naive Max-Sum, this approach is guaranteed to bound the true value if the ground truth matching is contained in the candidate set.

\vspace{0.5em} \noindent \textbf{Generalized Assignment Constrained (GA+C): } This approach applies our proposed algorithm with a constraint $N$ that corresponds to the 75\% percentile of the number of matches of every $r$ in the candidate set. Like the baseline Max-Sum-C, the result interval is no longer guaranteed to bound the true result but is often much more informative. 

\vspace{0.5em} \noindent \textbf{Generalized Assignment Constrained Optimal (GA*): } This approach applies our proposed algorithm with a constraint $N$ that corresponds to the best possible choice of constraint for the query, i.e., the constraint value that achieves the tightest bound that actually bounds the true result.

\begin{figure*}
    \centering
    \includegraphics[width=0.33\textwidth]{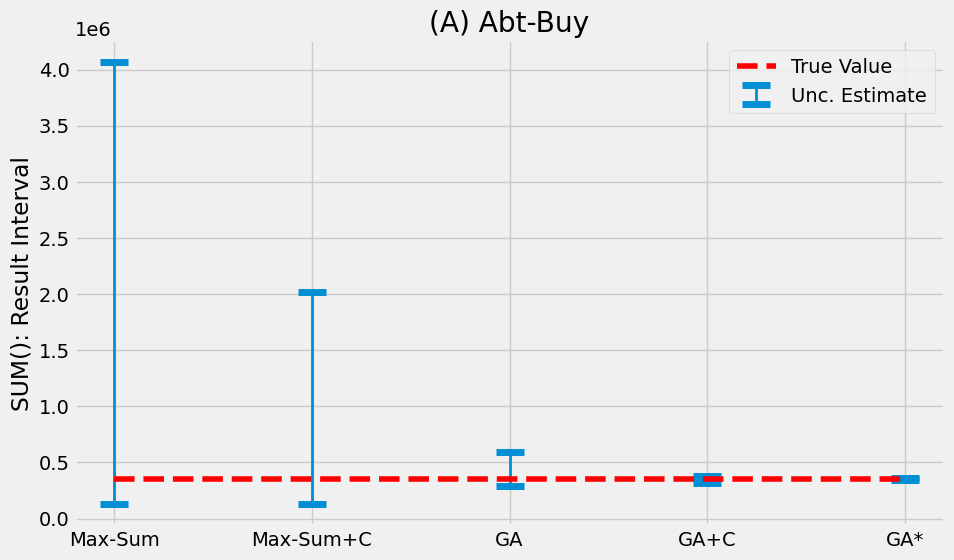}
    \includegraphics[width=0.33\textwidth]{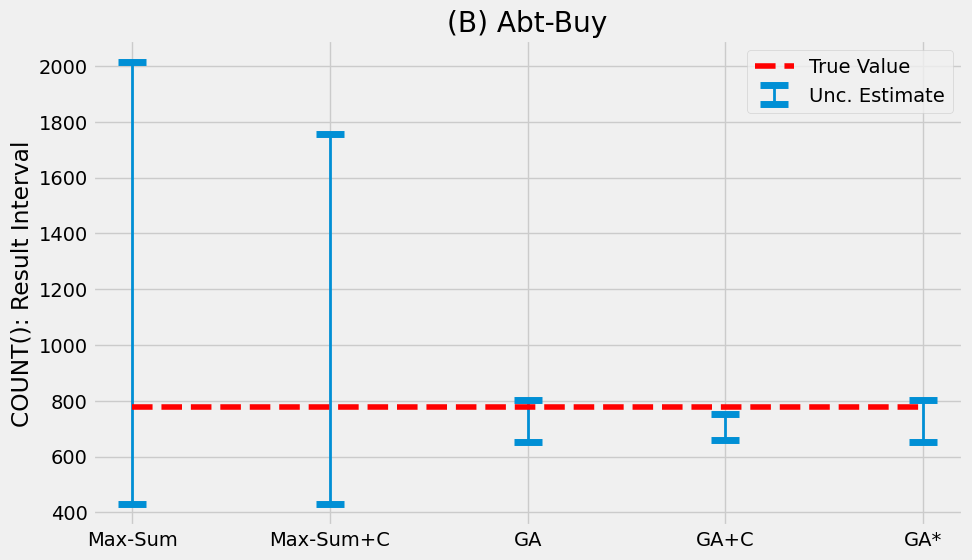}
    \includegraphics[width=0.33\textwidth]{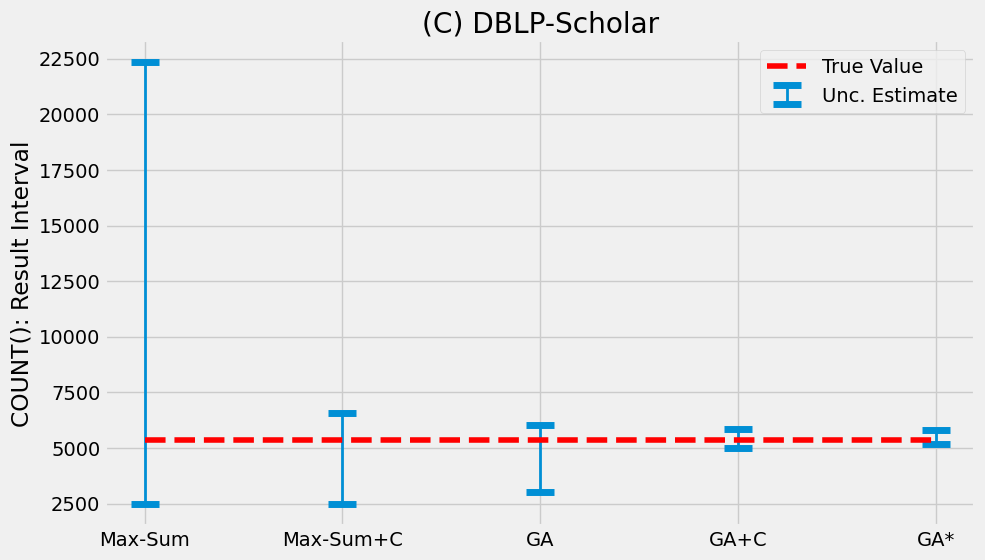}
    \includegraphics[width=0.33\textwidth]{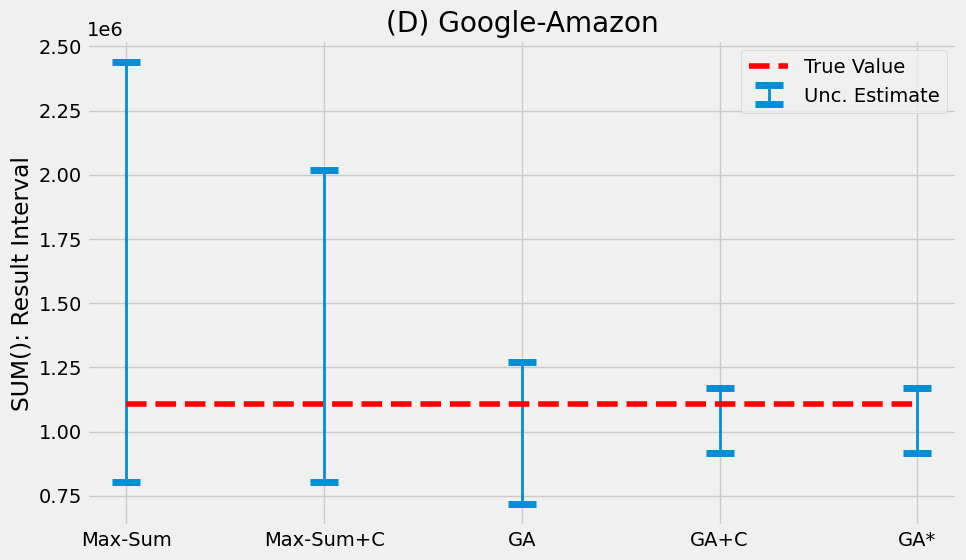}
    \includegraphics[width=0.33\textwidth]{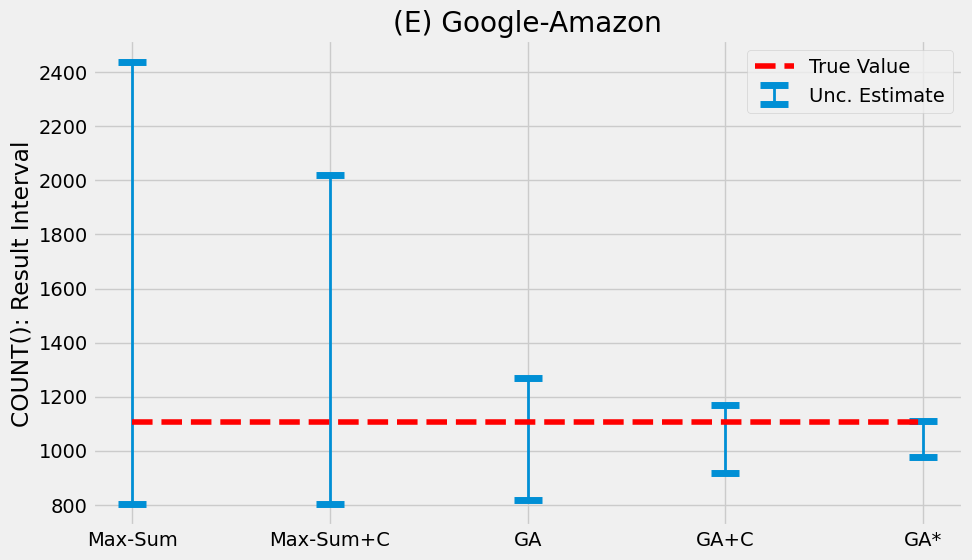}
    \includegraphics[width=0.33\textwidth]{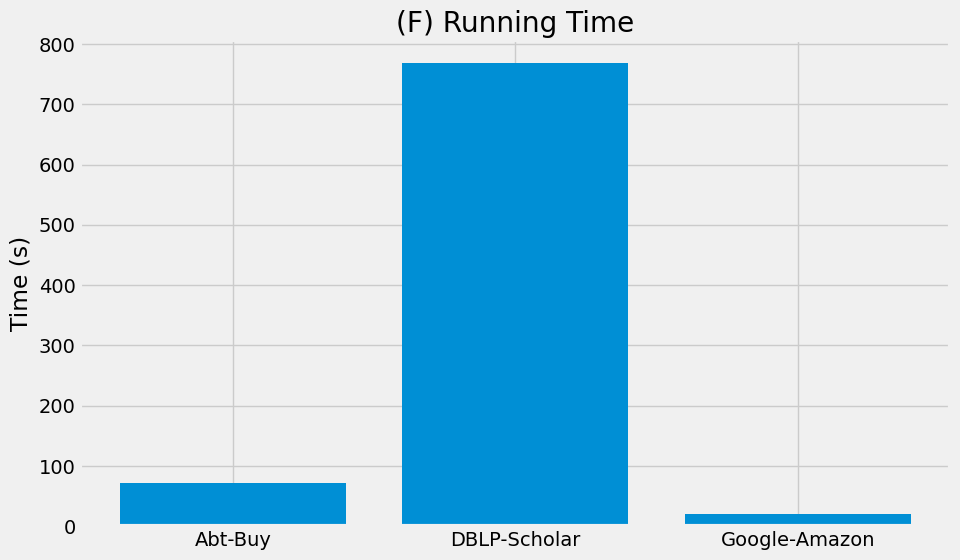}
    \caption{(A-E) Plots of the result intervals for the queries described in Section \ref{sec:query}. Our approaches provide significantly tighter bounds than the naive Max-Sum baseline. (F) The generalized assignment optimization problem can scale to real-world matching problem completing in minutes.}
    \label{fig:exp1}
\end{figure*}

\subsection{End-to-End Experiments on Real Data}
In our first set of experiments, we explore result interval calculation on the real datasets.

\subsubsection{Bounding Prototypical Queries}\label{sec:query}
In our first experiment, we consider 5 different prototypical queries over the three datasets. 
\begin{lstlisting}
qA = SELECT SUM(PRICE) FROM Abt_Buy 
qB = SELECT COUNT(TITLE) FROM Abt_Buy
qC = SELECT COUNT(TITLE) FROM DBLP_Scholar
qD = SELECT SUM(PRICE) FROM Amazon_Google
qE = SELECT COUNT(TITLE) FROM Amazon_Google
\end{lstlisting}
Figure \ref{fig:exp1} illustrates how our approach bounds the 5 different prototypical aggregate queries. 
The absolute ranges returned by each estimation technique are shown, and the true value is shown in red. 
At a high level, these plot illustrates how the approach works in real-world terms, and how much tighter the returned result intervals are compared to the Max-Sum baselines.
While these are five individual queries (and are, admittedly, picked to be illustrative), we will show in later experiments how the techniques perform over an entire workload.

Nonetheless, Figure \ref{fig:exp1}A-E shows that the Max-Sum result intervals are often too wide to be useful.
A direct application of~\cite{potti2015daq} will not work in this problem setting. In part, this is due to the quadratic nature of matching, where one $r$ could have an inopportune candidate set where it matches with nearly all $S$ rows. In all of the queries, the constrained versions of the result interval estimators far more tightly bound the true values. Furthermore, the heuristic choice of the 75\% percentile constrain works effectively to tighten the confidence intervals.
Figure \ref{fig:exp1}F, shows that the generalized assignment optimization problem can scale to real-world matching problem completing in minutes.

There are a number of general points about the problem that are also illustrated in the plots.
Firstly, the true value is not necessarily at the midpoint of the range.
This is why the tightness matters so much as a metric.
Second, the heuristic GA+C is generally close to the best choice of constraint GA* for these queries.

\begin{figure*}
    \centering
    \includegraphics[width=0.32\textwidth]{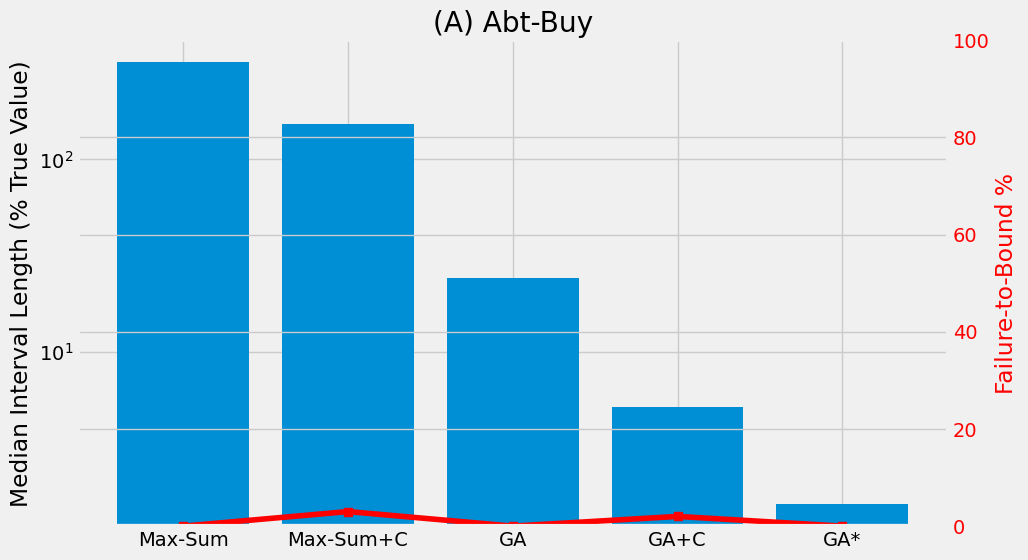}
    \includegraphics[width=0.32\textwidth]{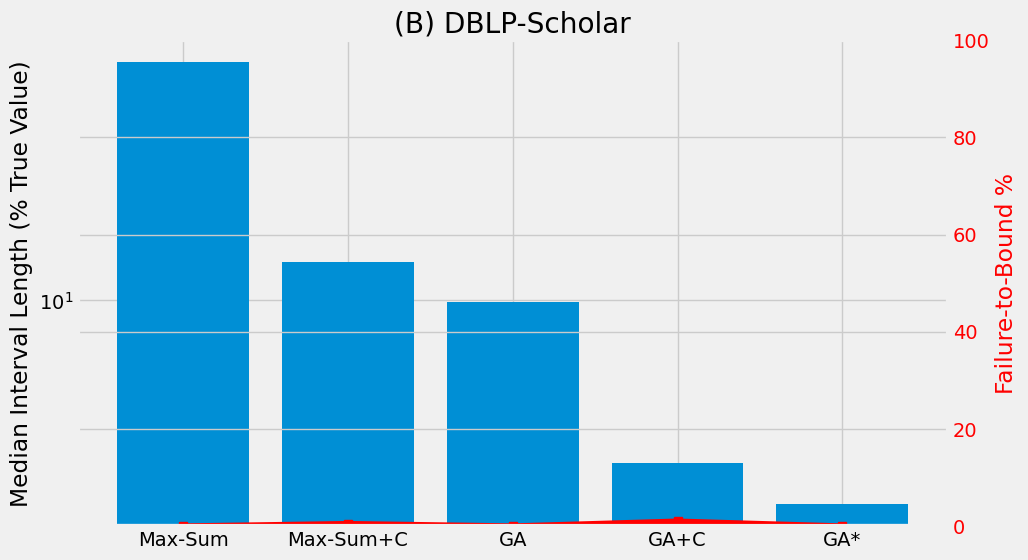}
    \includegraphics[width=0.32\textwidth]{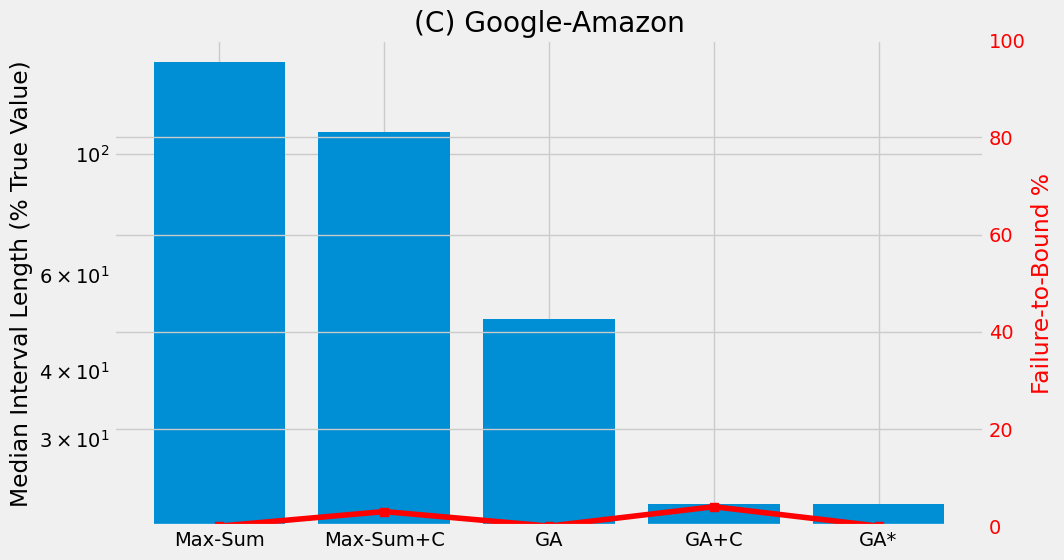}
    \caption{(A-C) The result interval tightness and their reliability on the three real datasets.}
    \label{fig:exp2}
\end{figure*}

\subsubsection{Overall Accuracy Analysis}
The last section presented results only on hand-picked queries. These are illustrative to understand how the estimation works and behaves. Now, we present a more comprehensive analysis of accuracy on randomly generated queries for each dataset. These queries are like the prototypical ones above but have random predicates. 
\begin{lstlisting}
qA = SELECT SUM(Price) FROM Abt_Buy WHERE <random key word in title>
qB = SELECT COUNT(1) FROM DBLP_Goog WHERE <random year>
qD = SELECT SUM(Price) FROM Goog_Amzn WHERE <random key word in title>
\end{lstlisting}
Figure \ref{fig:exp2} illustrates the results. We plot these results on two axes. First, we show the result interval length (relative to the true value). Then, we show the failure-to-bound rate (the fraction of true results that were outside the interval). In an ideal scenario, the interval length should be small and the failure-to-bound rate should be zero. The optimal constraint value GA* is optimized for each individual query using ground truth (thus is hypothetical).
These results show that GA+C strikes the best balance between the result tightness and the failure-to-bound rate, which was less than 5\% of the queries.

\begin{figure*}[t]
    \centering
    \includegraphics[width=0.32\textwidth]{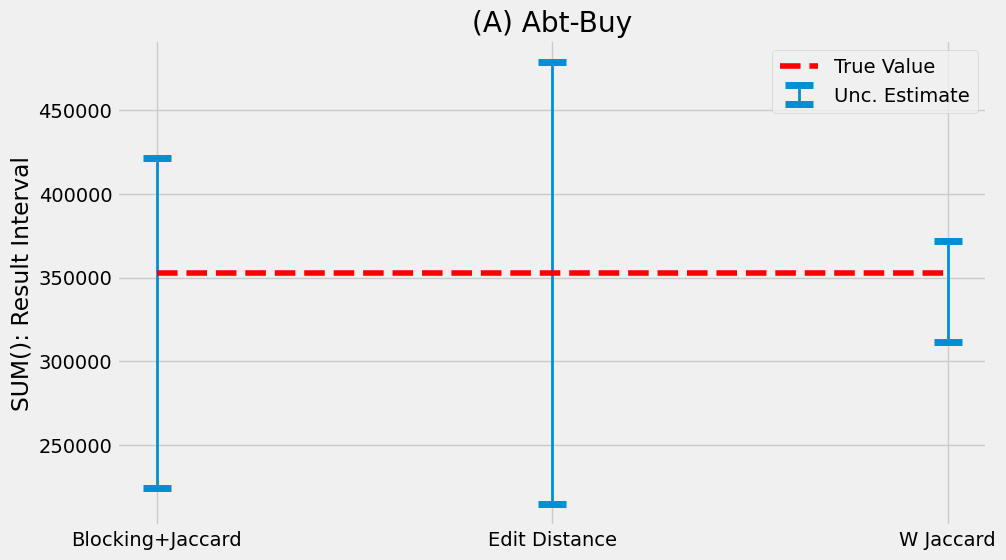}
    \includegraphics[width=0.32\textwidth]{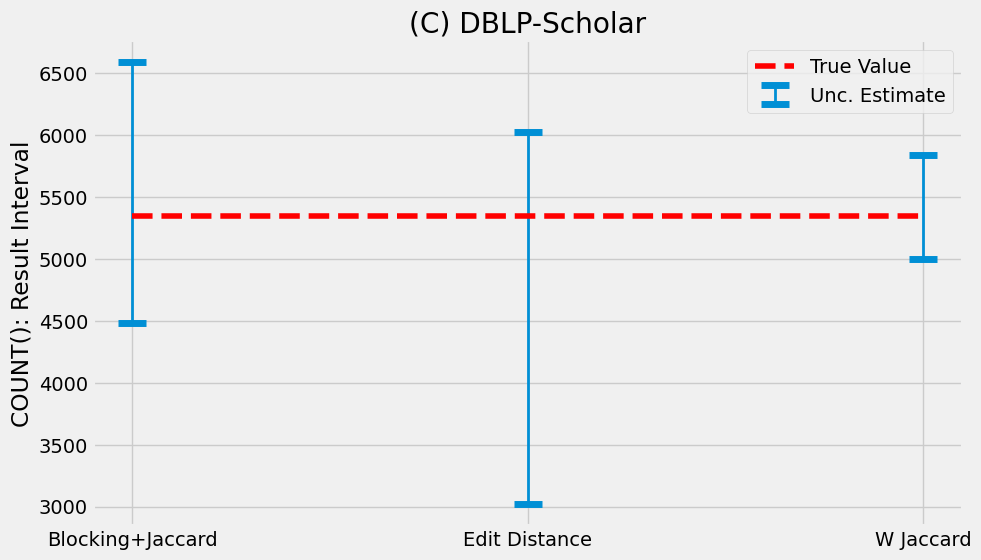}
    \includegraphics[width=0.32\textwidth]{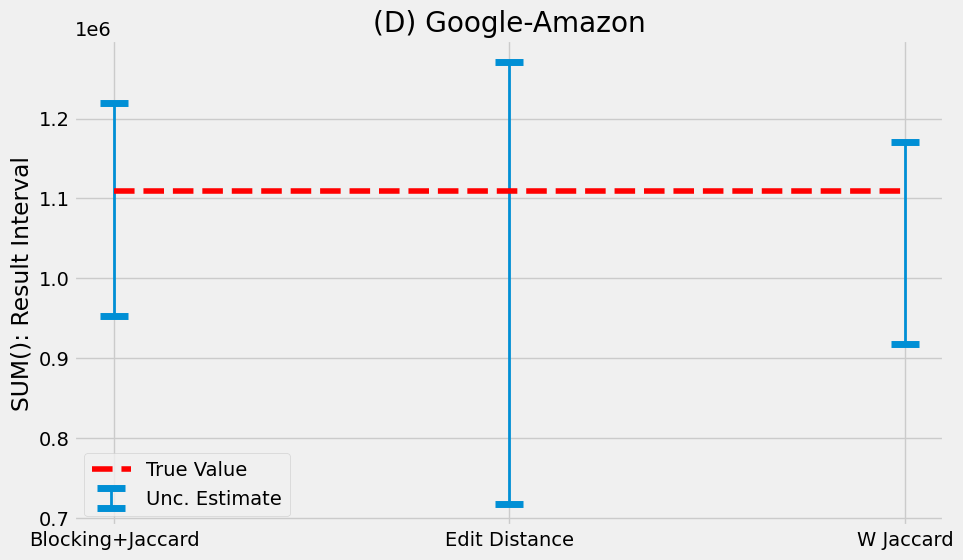}
    \caption{(A,C,D) We revisit three of the queries in the first experiment and present results with different candidate sets. The result intervals can give users guidance on how precise a record linkage strategy is in terms of how it affects aggregate queries. }
    \label{fig:exp3}
\end{figure*}

\subsubsection{Varying Data Integration Pipelines}
In the next experiment, we revisit the prototypical queries used in the first experiment.
We present results on different data integration pipelines, which manifests itself as different candidate sets for our GA algorithm.
In Figure \ref{fig:exp3} (labeled the same way as the first experiment), we vary the construction of the candidate set with different data integration pipelines. 
We present the result interval for the GA+C approach on three of the initial prototypical queries. We consider the following candidate sets.

\begin{itemize}
    \item \textbf{Blocking + Jaccard. } This approach first applies a blocking step to partition the dataset and then considers a Jaccard similarity comparison within the blocks. Abt-Buy and the Google-Amazon dataset are blocked on ``Price'', and the DBLP-Google dataset is blocked on ``Year''. Within each block the same Jaccard similarity threshold as before is used.
    
    \item \textbf{Edit Distance. } Instead of using the Jaccard similarity, we use an edit distance metric. We tune the similarity threshold until the candidate set size is roughly the same as the Jaccard threshold used before.
    
    \item \textbf{Weighted Jaccard. } We use a weighted Jaccard metric where tokens are weighted by their inverse-document frequency.
\end{itemize}

We chose these data integration pipelines to illustrate the calibration of our result interval estimates. It should be true that a less suitable metric has a wide interval. We show that this is the case in our data. It is known on these datasets that the Weighted Jaccard metric produces an effective candidate set. Consequently, the interval length is the smallest. Similarly,  edit distance is a poor choice for the title comparisons in these datasets.
This makes the edit distance interval the longest.
The blocking approach lies somewhere in the middle.
Users can use these intervals to understand how good or how bad different matching methodologies are across a dataset in terms of their impact on aggregate queries.

\begin{figure}
    \centering
    \includegraphics[width=0.49\columnwidth]{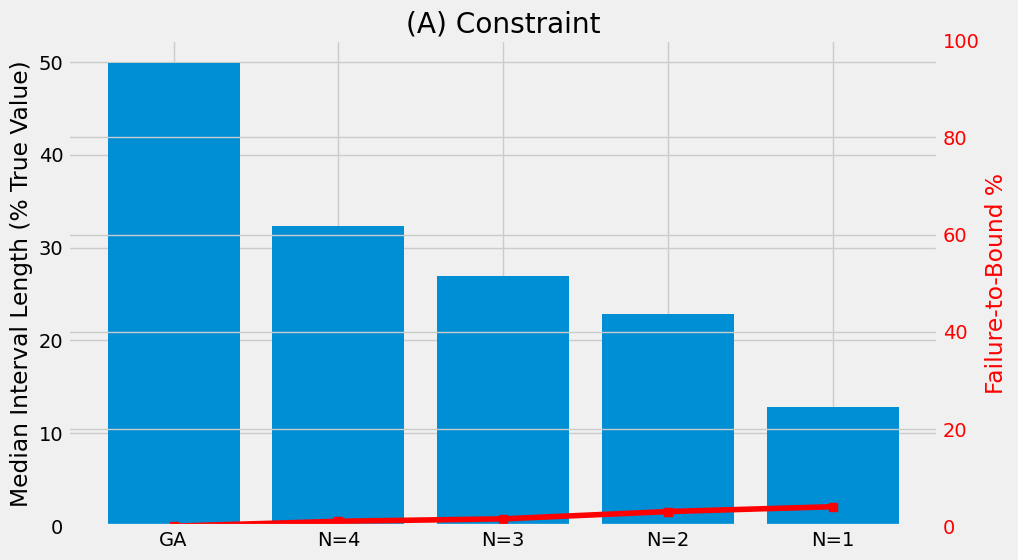}
    \includegraphics[width=0.49\columnwidth]{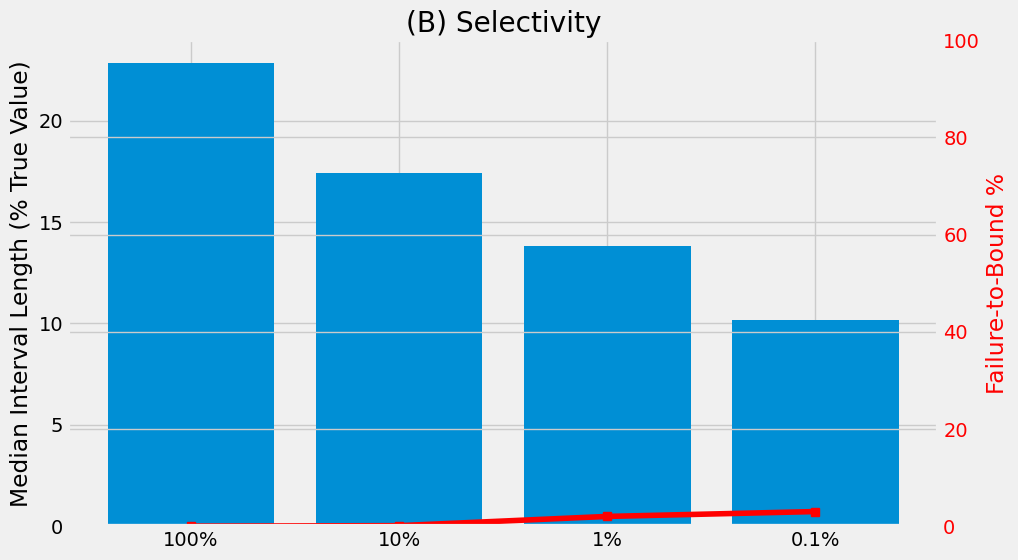}
    \caption{(A) Result interval tightness and failure-to-bound rate as a function of the constraint value, (B) Query selectivity and result interval tightness.}
    \label{fig:exp4}
\end{figure}

\subsection{Selectivity and Accuracy}
Next, we dig deeper into the relationship between the result interval length and the failure-to-bound rate. The GA algorithm is guaranteed to bound the result if the candidate set contains the ground truth matching. This is often hard to achieve in practice without making a permissive candidate set that errs on the side of false positives. On the other hand, such false positives can easily lead to extreme result interval estimates. 

The matching constraint allows one to control the sensitivity to extreme results. 
However, with the constraint, we are no longer guaranteed to bound the true result.
Intuitively, the tighter the intervals the more likely a failure happens. 
Figure \ref{fig:exp4}A illustrates in the Google-Amazon product matching case for the SUM query. As, we make the matching constraint stricter the interval size drops but the failure-to-bound rate increases. This relationship is not unlike those in statistical confidence intervals (e.g., $\pm 10$ with $95\%$ probability).

There is also an interesting relationship when one considers how query selectivity affects these metrics. 
Figure \ref{fig:exp4} plots the result interval length and the failure to bound rate as a function of query selectivity.
We apply predicates of different average selectivity to the Google-Amazon SUM query.
As the queries become more selective, the failure-to-bound rate increases.
This is because no candidate set is perfect, and those errors can get cancelled out for less selective queries.

\begin{figure}
    \centering
    \includegraphics[width=0.49\columnwidth]{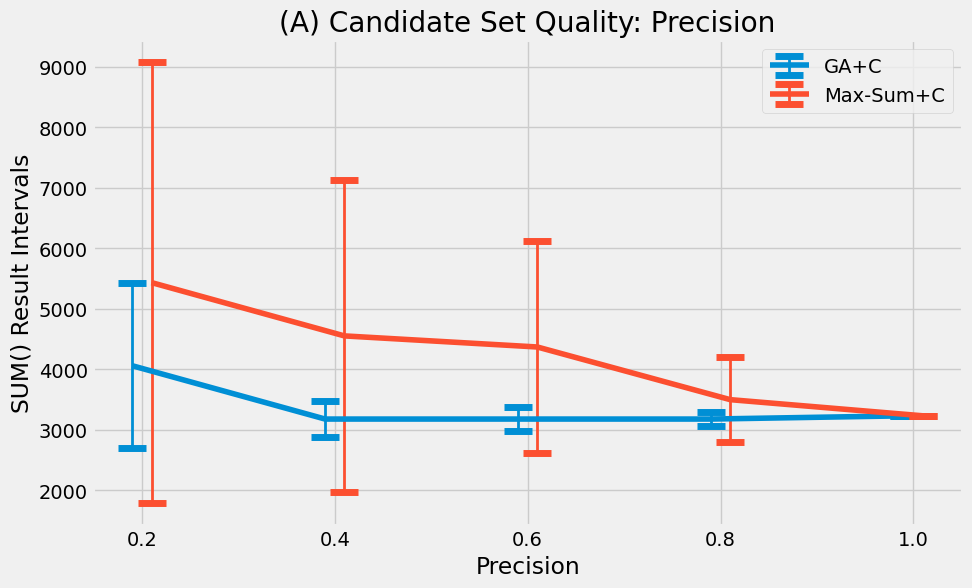}
    \includegraphics[width=0.49\columnwidth]{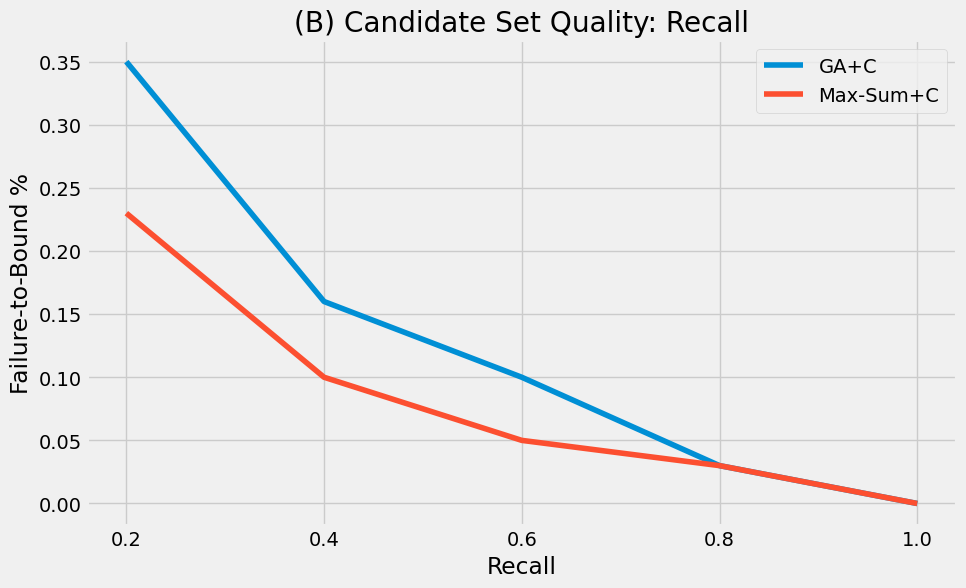}
    \caption{(A) How the precision of the candidate set affects result interval length, (B) How the recall of the candidate set affects the failure to bound rate.}
    \label{fig:exp5}
\end{figure}

\subsection{Synthetic Data Experiments}
Since ground-truth data is very difficult to find in a real-life setting for 1 to n matches, using synthetic datasets proved to be the best way to micro-benchmark the approach. 
The synthetic datasets are created using the following approach: first, the user provides a base table.  Then, the provided table is used as a baseline to create the second table. In the creation process of the second table, random typos are added with varying degrees (the range varies between Levenshtein distance of 1 to 3). 
Therefore, this dataset is generated with a known ground-truth matching and known similarity metric that relates the entities in both tables.

In order to reflect the real-life challenge of having a range of possible matchings for each distinct entry, the experiments are done with both balanced and skewed matching settings. In this context, balanced matching refers to a ground truth matching of exactly $n$ number of matches guarantee. Skewed matching refers to a 1 through $n$ number of possible matches for each entry. For the skewed matching setting, a randomly generated number between $1$ and $n$ is used in order to provide a randomized skew in the number of possible matches for each entry in the base table.
Over this dataset, we run SUM queries with randomly generated predicates uniformly with a selectivity of 0.1.

\subsubsection{Candidate Set Accuracy}
The first question that we can answer with this setup is how the accuracy of the candidate set affects the results. 
Figure \ref{fig:exp5} shows how different quality candidate sets affect results.
We generate the perfect mapping and incrementally add noise to the candidate set (either false positives in Figure \ref{fig:exp5}A or false negatives in Figure \ref{fig:exp5}B).

Figure \ref{fig:exp5}A shows how false positives affect the length of the result interval. A less precise candidate set will generally have wider intervals than a more precise one. 
This trend is observed in both the Max-Sum algorithm and the GA algorithm. We find that the GA algorithm is more robust to low-precision candidate sets.
On the other hand, Figure \ref{fig:exp5}B shows how false negatives in the candidate set affect the failure-to-bound rate.
Here, we do identify a weakness of our proposed algorithm where the optimization problem is sensitive to false negatives, or low recall candidate sets. 

\begin{figure}
    \centering
    \includegraphics[width=0.7\columnwidth]{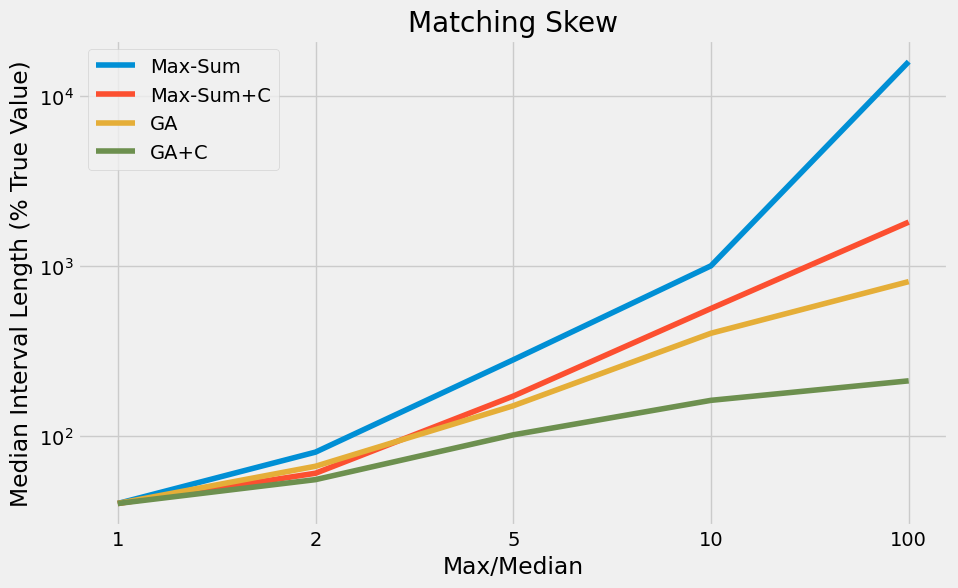}
    \caption{Skew in the candidate set is a key factor that governs how tight the result intervals are.}
    \label{fig:exp6}
\end{figure}

\subsubsection{Matching Skew}
Next, we consider the effect of skew in the matching set. We define skew as the maximum number of matches any individual base table row has in the candidate set divided by the median number of matches. 
Figure \ref{fig:exp6} plots the result interval length for each of the approaches.
It is clear that the Max-Sum approaches are highly sensitive to skewed data.
This is because of the double-counting issues that we talked about earlier in the paper.
The GA algorithm is far more effective at dealing with skew, especially with the match cardinality constraint.

\begin{figure}
    \centering
    \includegraphics[width=\columnwidth]{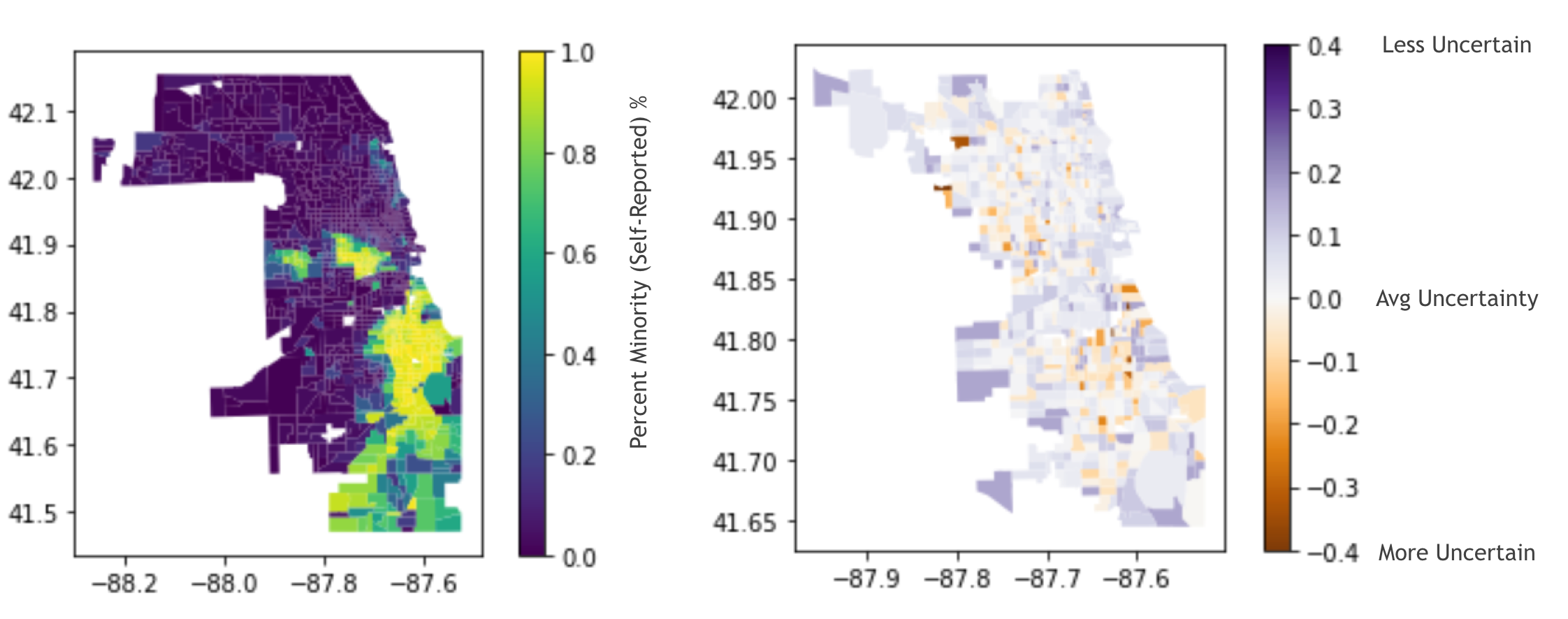}
    \caption{A real-world application of our framework to understand inequities in data quality in public data.}
    \label{fig:exp7}
\end{figure}

\subsection{Case Study: Produce Availability Survey}
Finally, the whole purpose of the proposed framework is to allow data scientists to assess uncertainty in data integration problems. We present a case study to illustrate the types of analysis that our framework would allow.

Many public health organizations recognize that, beyond
focusing on and treating biological mechanisms of disease, advancing health also critically requires accounting for and striving to mitigate adverse consequences of social, environmental, behavioral, and psychological factors. To date, such factors have not been comprehensively codified and quantified in a way suitable for large-scale co-analysis/data-mining with explicitly biological or clinical data to learn new insights into factors influencing wellness or disease. We worked with researchers at the University of Chicago Medical School to organize a pilot dataset that links patient data with social factors based on GPS location data that indicates key lifestyle factors. 

One such important factor is the availability of fresh food and produce near a patient's home~\cite{shannon2014food}. 
The City of Chicago maintains a dataset of all business licenses in the city~\footnote{https://data.cityofchicago.org/Community-Economic-Development/Business-Licenses-Current-Active/uupf-x98q}.
These licenses contain business activity descriptions so that one could identify stores that sell fresh food. However, these city-level classifications are not always precise. 

Consider two different establishments in the dataset that sell fresh produce. One of them is tagged appropriately and the other is not.
\begin{lstlisting}
South Loop Market,1720 S MICHIGAN AVE 1ST 115,CHICAGO,IL,60616.0,Retail Food Establishment, Retail Sales of Perishable Foods

Mariano's #8515,1800 W LAWRENCE AVE 1 & 2,CHICAGO,IL,60640,Retail Food Establishment,"Retail Sales of Perishable Foods | Retail Sales of Fresh Fruits and Vegetables | Preparation of Food, Coffee or Drinks | Deli, Butcher or Bakery"
\end{lstlisting}
Reconciling each one of these ambiguities by hand is extremely time-consuming. 
Luckily a prior dataset from 2013 exists that took a survey of such stores in Chicago~\cite{https://data.cityofchicago.org/Community-Economic-Development/Map-of-Grocery-Stores-2013/ce29-twzt}.
If an existing business is not appropriately tagged as a grocery store that sells fresh produce and it is contained in the old dataset, it is possibly a grocery store. 
Of course, the names and addresses from these two datasets do not completely align because they were collected in different time-periods. 

The combination of these two datasets will be heuristic, but 
we can apply our proposed GA optimization framework to determine our confidence in that merging process. Here's how it works:
\begin{itemize}
    \item \textbf{Base Table. } Current dataset of Chicago Business Listings that are not already tagged as grocery stores.
    \item \text{Augmenting Table. } Prior dataset of Grocery Store Listings in Chicago.
    \item \textbf{Candidate Set. } Use a Jaccard Similarity Matching over store name and address with threshold $0.7$.
    \item \textbf{Matching Cardinality Constraint. } Match with at most 1 base table entity.
    \item \textbf{Query. } Count the fresh fruit and produce stores in each census tract.
\end{itemize}

The result of this framework determines a range of grocery store counts per census tract. Intuitively, it gives a high estimate based on generous matchings between the augmenting table and the base table, and a low estimate by assuming they don't match. 

We can use this calculation to understand where there is uncertainty in this combined dataset and what kind of biases this uncertainty might introduce.
Some census tracts will have cleaner data, and others will have more ambigious data.
"Cleaner" can either mean that the current business listings are appropriately tagged or that there are clearer matches in the prior dataset.
Figure \ref{fig:exp7} illustrates the uncertainty calculations by our framework mapped across the census tracts in the city of Chicago. In fact, we found that the data quality issues were more severe in minority neighborhoods leading to more ambiguous estimates.

\makeatletter
\newcounter{savesection}
\newcounter{apdxsection}
\renewcommand\appendix{\par
  \setcounter{savesection}{\value{section}}%
  \setcounter{section}{\value{apdxsection}}%
  \setcounter{subsection}{0}%
  \gdef\thesection{\@Alph\c@section}}
\newcommand\unappendix{\par
  \setcounter{apdxsection}{\value{section}}%
  \setcounter{section}{\value{savesection}}%
  \setcounter{subsection}{0}%
  \gdef\thesection{\@arabic\c@section}}
\makeatother
\section{Conclusion}
To conclude, this paper formalizes a measure of uncertainty in two-table, one-to-many data integration workflows.
Such a measure can help users understand how data integration choices can affect downstream aggregate query processing. 
We propose an algorithmic framework based on graph matching to efficiently calculate this uncertainty measure for different SQL predicates and aggregation functions of interest. Finally, we illustrate how these uncertainty metrics can be used to inform downstream data science applications in a real-world case study.

\appendix
\section{Appendix}
\subsection{Proof of AVG Query Bounds}
The key insight is that the following equality holds at both the attainment of the minimum and maximum:
\[
\forall s \in S: \sum_{r \in R} x_{r,s} = 1
\]
That is, that every $s \in S$ is matched to one $R$.

Leveraging this insight, let's first consider bounding the minimum. 
Notice the denominator of the objective function above:
\[
\sum_{(r,s) \in \Psi} \textbf{sgn}(W(r,s)) \cdot x_{r,s}
\]
Th term summation is $\textbf{sgn}(W(r,s)) \cdot x_{r,s}$, which is the product of two binary variables. Both of these have to be equal to $1$ to increase the sum. Therefore,
\[
\sum_{(r,s) \in \Psi} \textbf{sgn}(W(r,s)) \cdot x_{r,s} \le \sum_{(r,s) \in \Psi} x_{r,s}
\]
We know both of the following expressions hold at the minimum:
\[
\forall r \in R: \sum_{r \in R} x_{r,s} \le N, 
\forall s \in S: \sum_{r \in R} x_{r,s} = 1
\]
Which means that:
\[
\sum_{(r,s) \in \Psi} \textbf{sgn}(W(r,s)) \cdot x_{r,s} \le \min\{|R|N,|S|\}
\]
It follows that this inequality holds for any minimizer of the SUM objective:
\[
\frac{\sum_{(r,s) \in \Psi} W(r,s) \cdot x_{r,s}}{\min\{|R|N,|S|\}} \le \frac{\sum_{(r,s) \in \Psi} W(r,s) \cdot x_{r,s}}{\sum_{(r,s) \in \Psi} \textbf{sgn}(W(r,s)) \cdot x_{r,s}} 
\]
Leading to:
\[
l_{avg} = \frac{1}{\min\{|R|N,|S|\}} \cdot l_{sum} \square
\]

Now, let's consider the maximum.
For each $s \in S$, let $d(s)$ be the following.
\[
d(s) = \max{r \in R} \textbf{sgn}(W(r,s))
\]
$d(s)$ is equal to $1$ if there exists at least one non-zero edge, and equal to zero otherwise. At attainment of the maximum sum,  the following must be true:
\[
\forall s \in S: \sum_{r \in R} x_{r,s} = 1
\]
It can be easily seen using set intersection logic that the following must hold :
\[
\sum_{(r,s) \in \Psi} \textbf{sgn}(W(r,s)) \cdot x_{r,s} \ge \sum_{s \in S} d(s)
\]
Which leads to the following inequality at every maximizer of the SUM for $d = \sum_{s \in S} d(s)$:
\[
\frac{\sum_{(r,s) \in \Psi} W(r,s) \cdot x_{r,s}}{d} \ge \frac{\sum_{(r,s) \in \Psi} W(r,s) \cdot x_{r,s}}{\sum_{(r,s) \in \Psi} \textbf{sgn}(W(r,s)) \cdot x_{r,s}} 
\]
And finally,
\[
u_{avg} = \frac{1}{d} \cdot u_{sum} \square
\]
\unappendix
\section{Acknowledgements}
In loving memory of Tahsin Türkçapar and Mehmet Fatih Taşgetiren, beloved family members and sources of unwavering support throughout this research journey. Their encouragement, values, and beliefs  were a constant source of motivation. This work stands as a testament to their enduring influence on people whose lives they've touched.

\bibliographystyle{abbrv}
\bibliography{refs}
\makeatletter
\let\ACM@origbaselinestretch\baselinestretch
\makeatother
\end{document}